\documentclass[aps, twocolumn, amsmath, superscriptaddress]{revtex4}
\usepackage{dcolumn}
\usepackage{bm}
\usepackage{graphicx}
\usepackage{epstopdf} 
\usepackage{color}
\usepackage{ulem}
\usepackage{amsfonts}
\DeclareGraphicsExtensions{. jpg,. pdf, . mps, . png, . eps, . ps, . EPS}

\usepackage[toc,page]{appendix}
\usepackage{adjustbox}
\usepackage{bm}

\usepackage{xcolor}
\usepackage{framed}
\definecolor{shadecolor}{rgb}{0.941406,0.78125,0.714844}

\begin{document}
\def\bc{\begin{center}} 
\def\ec{\end{center}}
\def\bea{\begin{eqnarray}}
\def\eea{\end{eqnarray}}
\newcommand{\norm}[1]{|{#1}|}
\newcommand{\avg}[1]{\langle{#1}\rangle}
\newcommand{\ket}[1]{\left |{#1}\right \rangle}
\newcommand{\beq}{\begin{equation}}
\newcommand{\eneq}{\end{equation}}
\newcommand{\beqnn}{\begin{equation*}}
\newcommand{\eneqnn}{\end{equation*}}
\newcommand{\beqy}{\begin{eqnarray}}
\newcommand{\eneqy}{\end{eqnarray}}
\newcommand{\beqynn}{\begin{eqnarray*}}
\newcommand{\eneqynn}{\end{eqnarray*}}
\newcommand{\half}{\mbox{$\textstyle \frac{1}{2}$}}
\newcommand{\proj}[1]{\ket{#1}\bra{#1}}
\newcommand{\av}[1]{\langle #1\rangle}
\newcommand{\braket}[2]{\langle #1 | #2\rangle}
\newcommand{\bra}[1]{\langle #1 | }
\newcommand{\Avg}[1]{\left\langle{#1}\right\rangle}
\newcommand{\inprod}[2]{\braket{#1}{#2}}
\newcommand{\upket}{\ket{\uparrow}}
\newcommand{\downket}{\ket{\downarrow}}
\newcommand{\Tr}{\mathrm{Tr}}
\newcommand{\hcontrol}{*!<0em,.025em>-=-{\Diamond}}
\newcommand{\hctrl}[1]{\hcontrol \qwx[#1] \qw}
\newenvironment{proof}[1][Proof]{\noindent\textbf{#1.} }{\ \rule{0.5em}{0.5em}}
\newtheorem{mytheorem}{Theorem}
\newtheorem{mylemma}{Lemma}
\newtheorem{mycorollary}{Corollary}
\newtheorem{myproposition}{Proposition}
\newcommand{\vp}{\vec{p}}
\newcommand{\Or}{\mathcal{O}}
\newcommand{\so}[1]{{\ignore{#1}}}

\newcommand{\cd}[1]{\textcolor{cyan}{CD: #1}}
\newcommand{\mb}[1]{\textcolor{green}{mb: #1}}
\newcommand{\ak}[2]{\textcolor{blue}{ak: #1}}

\newtheorem{thm}{Theorem}[section]
\newtheorem{lmm}[thm]{Lemma}
\newtheorem{cor}[thm]{Corollary}
\newtheorem{prop}[thm]{Proposition}
\newtheorem{defn}[thm]{Definition}
\newcommand{\argmax}{\operatorname{argmax}}
\newcommand{\argmin}{\operatorname{argmin}}
\newcommand{\bbb}{\mathbf{B}}
\newcommand{\bbg}{\mathbf{g}}
\newcommand{\bbr}{\mathbf{R}}
\newcommand{\bbw}{\mathbf{W}}
\newcommand{\bbx}{\mathbf{X}}
\newcommand{\bbxa}{\mathbf{X}^*}
\newcommand{\bbxb}{\mathbf{X}^{**}}
\newcommand{\bbxp}{\mathbf{X}^\prime}
\newcommand{\bbxpp}{\mathbf{X}^{\prime\prime}}
\newcommand{\bbxt}{\tilde{\mathbf{X}}}
\newcommand{\bby}{\mathbf{Y}}
\newcommand{\bbz}{\mathbf{Z}}
\newcommand{\bbzt}{\tilde{\mathbf{Z}}}
\newcommand{\bigavg}[1]{\biggl\langle #1 \biggr\rangle}
\newcommand{\bp}{b^\prime}
\newcommand{\bx}{\mathbf{x}}
\newcommand{\by}{\mathbf{y}}
\newcommand{\cc}{\mathbb{C}}
\newcommand{\cov}{\mathrm{Cov}}
\newcommand{\dd}{\mathcal{D}}
\newcommand{\ee}{\mathbb{E}}
\newcommand{\fp}{f^\prime}
\newcommand{\fpp}{f^{\prime\prime}}
\newcommand{\fppp}{f^{\prime\prime\prime}}
\newcommand{\ii}{\mathbb{I}}
\newcommand{\gp}{g^\prime}
\newcommand{\gpp}{g^{\prime\prime}}
\newcommand{\gppp}{g^{\prime\prime\prime}}
\newcommand{\hess}{\operatorname{Hess}}
\newcommand{\ma}{\mathcal{A}}
\newcommand{\mf}{\mathcal{F}}
\newcommand{\mi}{\mathcal{I}}
\newcommand{\ml}{\mathcal{L}}
\newcommand{\cp}{\mathcal{P}}
\newcommand{\mx}{\mathcal{X}}
\newcommand{\mxp}{\mathcal{X}^\prime}
\newcommand{\mxpp}{\mathcal{X}^{\prime\prime}}
\newcommand{\my}{\mathcal{Y}}
\newcommand{\myp}{\mathcal{Y}^\prime}
\newcommand{\mypp}{\mathcal{Y}^{\prime\prime}}
\newcommand{\pp}{\mathbb{P}}
\newcommand{\ppr}{p^\prime}
\newcommand{\pppr}{p^{\prime\prime}}
\newcommand{\ra}{\rightarrow}
\newcommand{\rr}{\mathbb{R}}
\newcommand{\smallavg}[1]{\langle #1 \rangle}
\newcommand{\sss}{\sigma^\prime}
\newcommand{\st}{\sqrt{t}}
\newcommand{\sst}{\sqrt{1-t}}
\newcommand{\tr}{\operatorname{Tr}}
\newcommand{\uu}{\mathcal{U}}
\newcommand{\var}{\mathrm{Var}}
\newcommand{\ve}{\varepsilon}
\newcommand{\vpp}{\varphi^{\prime\prime}}
\newcommand{\ww}{W^\prime}
\newcommand{\xp}{X^\prime}
\newcommand{\xpp}{X^{\prime\prime}}
\newcommand{\xt}{\tilde{X}}
\newcommand{\xx}{\mathcal{X}}
\newcommand{\yp}{Y^\prime}
\newcommand{\ypp}{Y^{\prime\prime}}
\newcommand{\zt}{\tilde{Z}}
\newcommand{\zz}{\mathbb{Z}}
 
\newcommand{\fpar}[2]{\frac{\partial #1}{\partial #2}}
\newcommand{\spar}[2]{\frac{\partial^2 #1}{\partial #2^2}}
\newcommand{\mpar}[3]{\frac{\partial^2 #1}{\partial #2 \partial #3}}
\newcommand{\tpar}[2]{\frac{\partial^3 #1}{\partial #2^3}}

\newcommand{\est}{e^{-t}}
\newcommand{\esst}{\sqrt{1-e^{-2t}}}
\newcommand{\bbu}{\mathbf{u}}
\newcommand{\bbv}{\mathbf{v}}
\newcommand{\bos}{{\boldsymbol \sigma}}
\newcommand{\bz}{\mathbf{z}}
\newcommand{\bg}{\mathbf{g}}

\newcommand{\red}[1]{\textcolor{red}{#1}}
\newcommand{\blue}[1]{\textcolor{blue}{#1}}

\newcommand{\XXX}[3]{{\color{blue}{\bf [#1:} {#3} {\it -#2-}{\bf ]}}}

\setlength{\fboxrule}{5pt}

\title{The shape of shortest paths in random spatial networks}

\author{Alexander P. Kartun-Giles\thanks{alexanderkartungiles@gmail.com}}
\affiliation{Max Plank Institute for Mathematics in the Sciences, Inselstr. 22, Leipzig, Germany}
\author{Marc Barthelemy\thanks{marc.barthelemy@ipht.fr}}
\affiliation{Institut de Physique Th\'{e}orique, CEA, CNRS-URA 2306, Gif-sur-Yvette, France
}
\author{Carl P. Dettmann\thanks{carl.dettmann@bristol.ac.uk}}
\affiliation{School of Mathematics, University of Bristol, University Walk, Bristol BS8 1TW, UK}

\begin{abstract}

  In the classic model of first passage percolation, for pairs of
  vertices separated by a Euclidean distance $L$, geodesics exhibit
  deviations from their mean length $L$ that are of order $L^\chi$,
  while the transversal fluctuations, known as wandering, grow as
  $L^\xi$. We find that when weighting edges directly with their
  Euclidean span in various spatial network models, we have two
  distinct classes defined by different exponents $\xi=3/5$ and $\chi = 1/5$, or $\xi=7/10$ and
  $\chi = 2/5$, depending only on coarse details of the specific
  connectivity laws used. Also, the travel time fluctuations are
  Gaussian, rather than Tracy-Widom, which is rarely seen in first
  passage models. The first class contains
  proximity graphs such as the hard and soft random geometric graph,
  and the $k$-nearest neighbour random geometric graphs, where via
  Monte Carlo simulations we find $\xi=0.60\pm 0.01$ and
  $\chi = 0.20\pm 0.01$, showing a theoretical minimal wandering. The
  second class contains graphs based on excluded regions such as
  $\beta$-skeletons and the Delaunay triangulation and are
  characterized by the values $\xi=0.70\pm 0.01$ and
  $\chi = 0.40\pm 0.01$, with a nearly theoretically maximal wandering
  exponent.  We also show numerically that the KPZ relation
  $\chi = 2\xi -1$ is satisfied for all these models. These results
  shed some light on the Euclidean first passage process, but also
  raise some theoretical questions about the scaling laws and the
  derivation of the exponent values, and also whether a model can be
  constructed with maximal wandering, or non-Gaussian travel
  fluctuations, while embedded in space.

\end{abstract}

\maketitle


\section{Introduction}\label{sec:1}

Many complex systems assume the form of a spatial network
\cite{barthelemy2011,barthelemy2018}. Transport networks, neural
networks, communication and wireless sensor networks, power and energy
networks, and ecological interaction networks are all important
examples where the characteristics of a spatial network structure are
key to understanding the corresponding emergent dynamics.

Shortest paths form an important aspect of their study. Consider for
example the appearance of \textit{bottlenecks} impeding traffic flow
in a city \cite{barthelemy2013,kirkley2018}, the emergence of spatial
small worlds \cite{amaral2000,hirsch2019}, bounds on the diameter of
spatial preferential attachment graphs
\cite{flaxman2004,jordan2010,jacob2015}, the random connection model
\cite{kartungiles2018,kog105,knight2017,privault2019}, or in spatial
networks generally \cite{aldous2009,aldous2010}, as well as geometric
effects on betweenness centrality measures in complex networks
\cite{kog105,crucitti2006}, and navigability \cite{boguna2008}.

\textit{First passage percolation} (FPP) \cite{hammersley1965}
attempts to capture these features with a probabilistic model. As with
\textit{percolation} \cite{grimmettbook}, the effect of spatial
disorder is crucial. Compare this to the elementary \textit{random
  graph} \cite{erdos1959}. In FPP one usually considers a
deterministic lattice such as $\mathbb{Z}^{d}$ with independent,
identically distributed weights, known as \textit{local passage
  times}, on the edges. With a fluid flowing outward from a point, the
question is, what is the minimum passage time over all possible routes
between this and another distant point, where routing is quicker along
lower weighted edges?

More than 50 years of intensive study of FPP has been carried out
\cite{auffinger2017}. This has lead to many results such as the
subadditive ergodic theorem, a key tool in probability theory, but
also a number of insights in crystal and interface growth
\cite{takeuchi2010}, the statistical physics of traffic jams
\cite{grimmettbook}, and key ideas of universality and scale
invariance in the shape of shortest paths \cite{takeuchi2011}. As an
important intersection between probability and geometry, it is also
part of the mathematical aspects of a theory of gravity beyond general
relativity, and perhaps in the foundations of quantum mechanics, since
it displays fundamental links to complexity, emergent phenomena, and
randomness in physics \cite{berry1978,bianconi2016}.

A particular case of FPP is the topic of this article, known as
\textit{Euclidean first passage percolation} (EFPP). This is a
probabilistic model of fluid flow between points of a $d$-dimensional
Euclidean space, such as the surface of a hypersphere. One studies
optimal routes from a source node to each possible destination node in
a spatial network built either randomly or deterministically on the
points. Introduced by Howard and Newman much later in 1997
\cite{howard1997} and originally a weighted complete graph, we adopt a
different perspective by considering edge weights given
deterministically by the Euclidean distances between the spatial
points themselves. This is in sharp contrast with the usual FPP
problem, where weights are i.i.d. random variables.

\begin{figure}
     \includegraphics[scale=0.5]{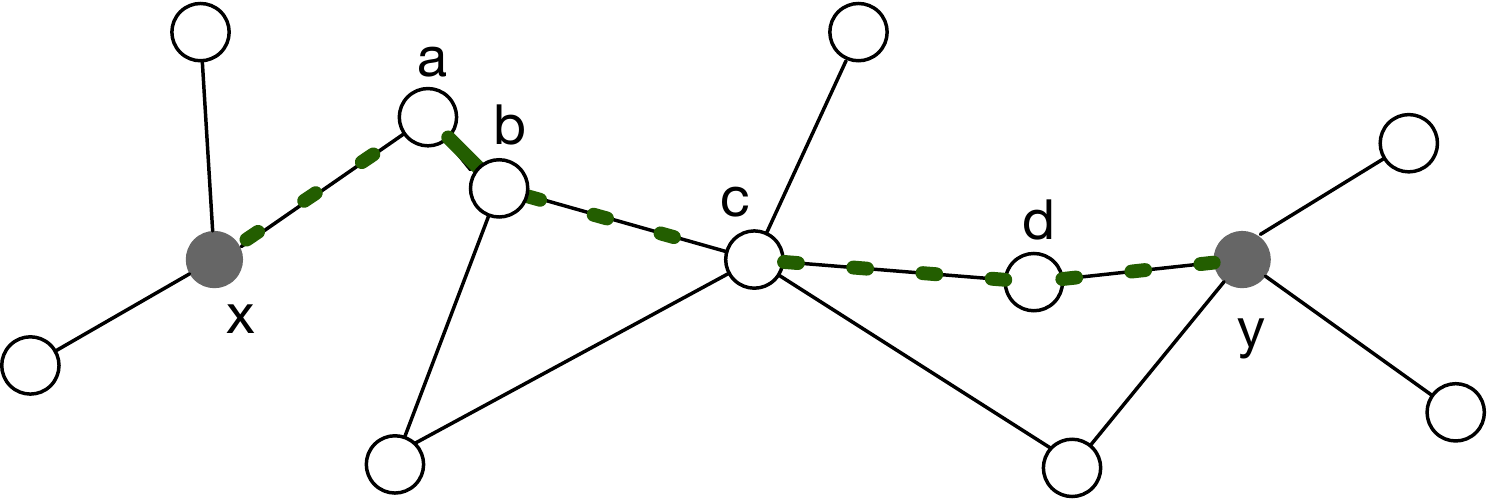}
     \caption{Illustration of the problem on a small network. The
       network is constructed over a set of points denoted by circles
       here and the edges are denoted by lines. For a pair of
       nodes $(x,y)$ we look for the shortest path (shown here by a
       dotted line) where the length of
       the path is given by the sum of all edges length:
       $d(x,y)=|x-a|+|a-b|+|b-c|+|c-d|+|d-y|$.}
 \label{fig:sp}
\end{figure}

Howard's model is defined on the complete graph constructed on a point
process. Long paths are discouraged by taking powers of interpoint
distances as edge weights. The variant of EFPP we study is instead
defined on a Poisson point process in an unbounded region (by
definition, the number of points in a bounded region is a Poisson
random variable, see for example \cite{chiubook}), but with links
added between pairs of points according to given rules
\cite{bollobas2001, penrosebook}, rather than the totality of the
weighted complete graph. More precisely, the model we study in this
paper is defined as
follows. We take a random spatial network such as the random geometric
graph constructed over a simple Poisson point process on a flat torus,
and weight the edges with their Euclidean length (see
Fig.~\ref{fig:sp}).
We then study the random length and transversal deviation of the shortest paths
between two nodes in the network, denoted $x$ and $y$, conditioned to
lie at mutual Euclidean separation $|x-y|$, as a function of the point
process density and other parameters of the model used (here and in
the following $|x|$ denotes the usual norm in euclidean space). The
study of the scaling with $|x-y|$ of the length and the deviation allow to define the fluctuation
and wandering exponents (see precise definitions below). We will
consider a variety of networks such as the random geometric graph with unit disk and Rayleigh fading
connection functions, the $k$-nearest neighbour graph, the Delaunay
triangulation, the relative neighbourhood graph, the Gabriel graph,
and the complete graph with (in this case only) the edge weights
raised to the power $\alpha>1$. We describe these models in more
detail in Section \ref{sec:spanet}.

To expand on two examples, the \textit{random geometric graph} (RGG)
is a spatial network in which links are made between all pair of
points with mutual separation up to a threshold.  This has
applications in e.g. wireless network theory, complex engineering
systems such as \textit{smart grid}, granular materials, neuroscience,
spatial statistics, topological data analyis, and complex networks 
\cite{last2017,penrose20162,hirsch2015,giles2016,kartungiles2018,kartungiles2019,kartungiles20192,ostilli2015,cunningham2019,
mulder2018}.

This paper is structured as follows. We first recap known results
obtained for both the FPP and Euclidean FPP in Section
\ref{sec:background}. We also discuss previous literature for the FPP
in non-typical settings such as random graphs and tessellations.
The reader eager to view the results can skip this section at first
reading, apart from the definitions of \ref{sec:background}A, however
the remaining background is very helpful for appreciating the later
discussion. In the Section \ref{sec:spanet} we introduce the various spatial networks
studied here, and in Section \ref{sec:results} we present the numerical
method and our new results on the EFPP model on random
graphs. In particular, due to arguments based on scale invariance, we
expect the appearance of power laws and universal exponents
\cite[Section 1]{takeuchi2011}. We reveal the scaling exponents of the
geodesics for the complete graph and for the network models
studied here, and also show numerical results about the travel time
and transversal deviation distribution. In particular, we find
distinct exponents from the KPZ class (see for example
\cite{halpin1995} and references therein) which has
wandering and fluctuation exponents $\xi=2/3$ and $\chi=1/3$,
respectively. Importantly, we conjecture and numerically corroborate a
Gaussian central limit theorem for the travel time fluctuations, on
the scale $t^{1/5}$ for the RGG and the other proximity graphs, and
$t^{2/5}$ for the Delaunay triangulation and other excluded region
graphs, which is also distinct from KPZ where the Tracy-Widom
distribution, and the scale $t^{1/3}$, is the famous outcome. Finally,
in Section \ref{sec:5} we present some analytic ideas
which help explain the distinction between universality classes. We
then conclude and discuss some open questions in Section \ref{sec:6}.

\section{Background: FPP and EFPP}
\label{sec:background}

\begin{figure*}
   
     \begin{adjustbox}{width=\linewidth}
     \centering
       \includegraphics[scale=0.4]{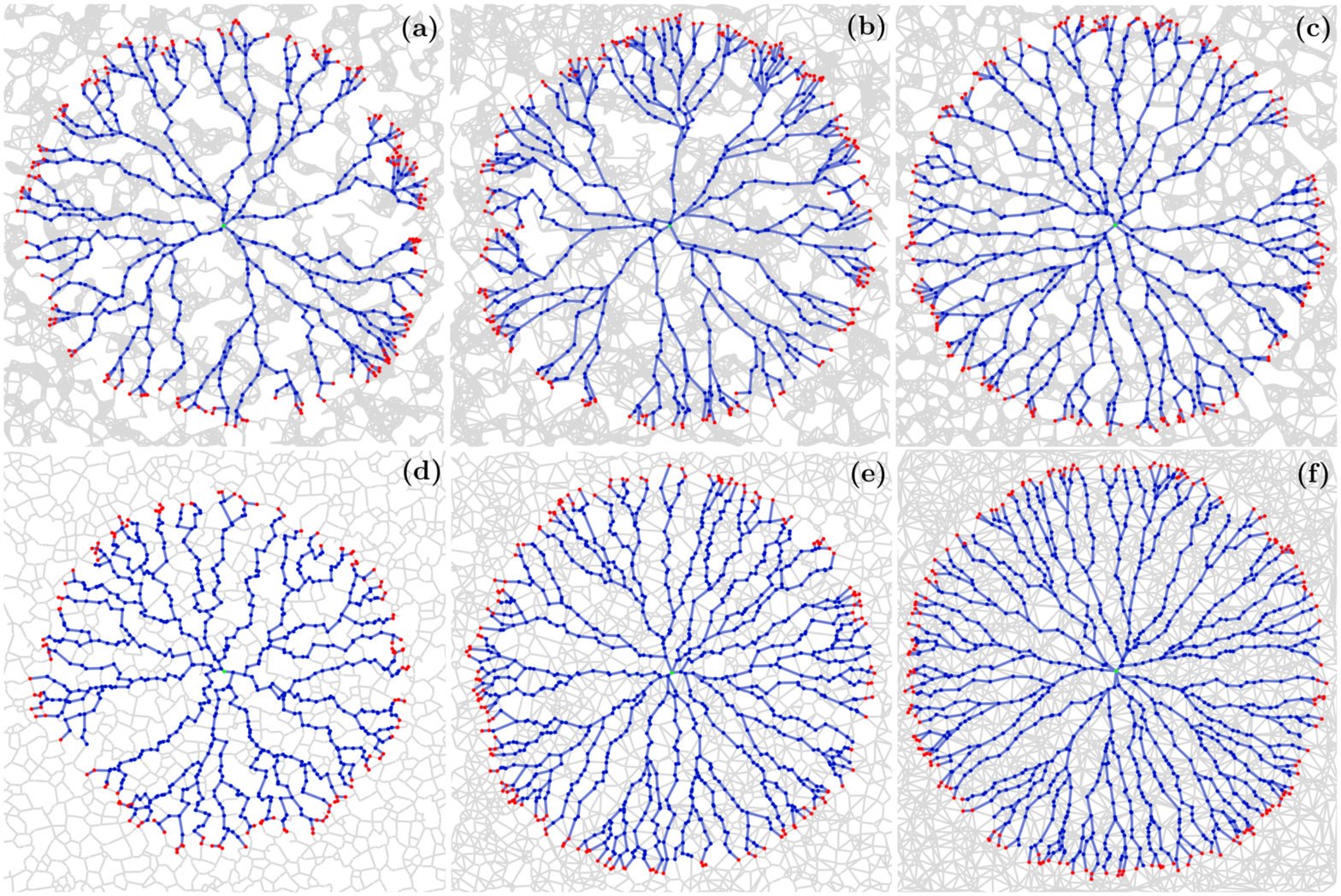}
        \end{adjustbox}

    \caption{Spatial networks, each built on a different realisation of a simple, stationary
      Poisson point processes of expected $\rho = 1000$ points in the
      unit square $\mathcal{V} = [-1/2,1/2]^2$, but with different connection laws. The boundary points at time $t=1/2$ of the first passage process are shown in red, while their respective geodesics are shown in blue. (a) Hard
      RGG with unit disk connectivity.  (b) Soft RGG
      with Rayleigh fading connection function
      $H(r)=\exp(-\beta r^2)$, (c) $7$-NNG, (d) Relative
     neighbourhood graph, which is the lune-based $\beta$-skeleton for $\beta=2$,
     (e) Gabriel graph, which is the lune-based $\beta$-skeleton for
     $\beta=1$, and (f) the Delaunay triangulation.}
    \label{fig:main}

\end{figure*}

In EFPP, we first construct a Poisson point process in $\mathbb{R}^d$
which forms the basis of
an undirected graph. A fluid or current then flows outward from a
single source at a constant speed with a travel time along an edge
given by a power $\alpha \geq 1$ of the Euclidean length of the edge
along which it travels \cite{howard1997}. See Fig. \ref{fig:main},
where the model is shown on six different random spatial network
models.

Euclidean FPP on a large family of connected random geometric graphs has been studied in detail by Hirsch, Neuh\"{a}user, Gloaguen and Schmidt in \cite{neuhauser2012,hirsch2015,hirsch20152} and the closely related works \cite{neuhauser2012,brereton2014,neuhauser2015,hirsch2015,hirsch2016,hirsch20152,coupier2018}, and references therein. Developing FPP in this setting, Santalla et al \cite{santalla2015}
recently studied the model on spatial networks, as we do here. Instead
of EFPP, they weight the edges of the Delaunay triangulation, and also
the square lattice, with i.i.d. variates, for example $\text{U}[a,b]$
for $a,b>0$, and proceed to numerically verify the existence of the
KPZ class for the geodesics, see e.g. \cite{chatterjee2013}, and the earlier work of Pimentel \cite{pimentel2011} giving the asymptotic first passage times for the Delaunay triangulation with i.i.d weights. Moreover,
FPP on small-world networks and Erd\H{o}s-Renyi random graphs are
studied by Bhamidi, van der Hofstad and Hooghiemstra in
\cite{bhamidi2010}, who discuss applications in diverse fields such as
magnetism \cite{abraham1995}, wireless ad hoc networks
\cite{beyme2014,knight2017,kartungiles2018}, competition in ecological
systems \cite{khordzhakia2005}, and molecular biology
\cite{bundschuh2000}. See also their work specifically on random
graphs \cite{hofstad2001}.  Optimal paths in disordered complex
networks, where disorder is weighting the edges with i.i.d. random
variables, is studied by Braunstein et al. in \cite{braunstein2003},
and later by Chen et al. in \cite{chen2006}. We also point to the recent
analytic results of Bakhtin and Wu, who have provided a good lower
bound rate of growth of geodesic wandering, which in fact we find to
be met with equality in the random geometric graph \cite{bakhtin2019}.

To highlight the difference between these results and our own, we have edge weights which are \textit{not}
 independent random variables, but interpoint distances. As far as we are aware, this has not been addressed directly in the literature.

 \subsection{First passage percolation} \label{sec:fpp}

 \begin{figure*}
   \begin{adjustbox}{width=\linewidth}
     \includegraphics[scale=0.292]{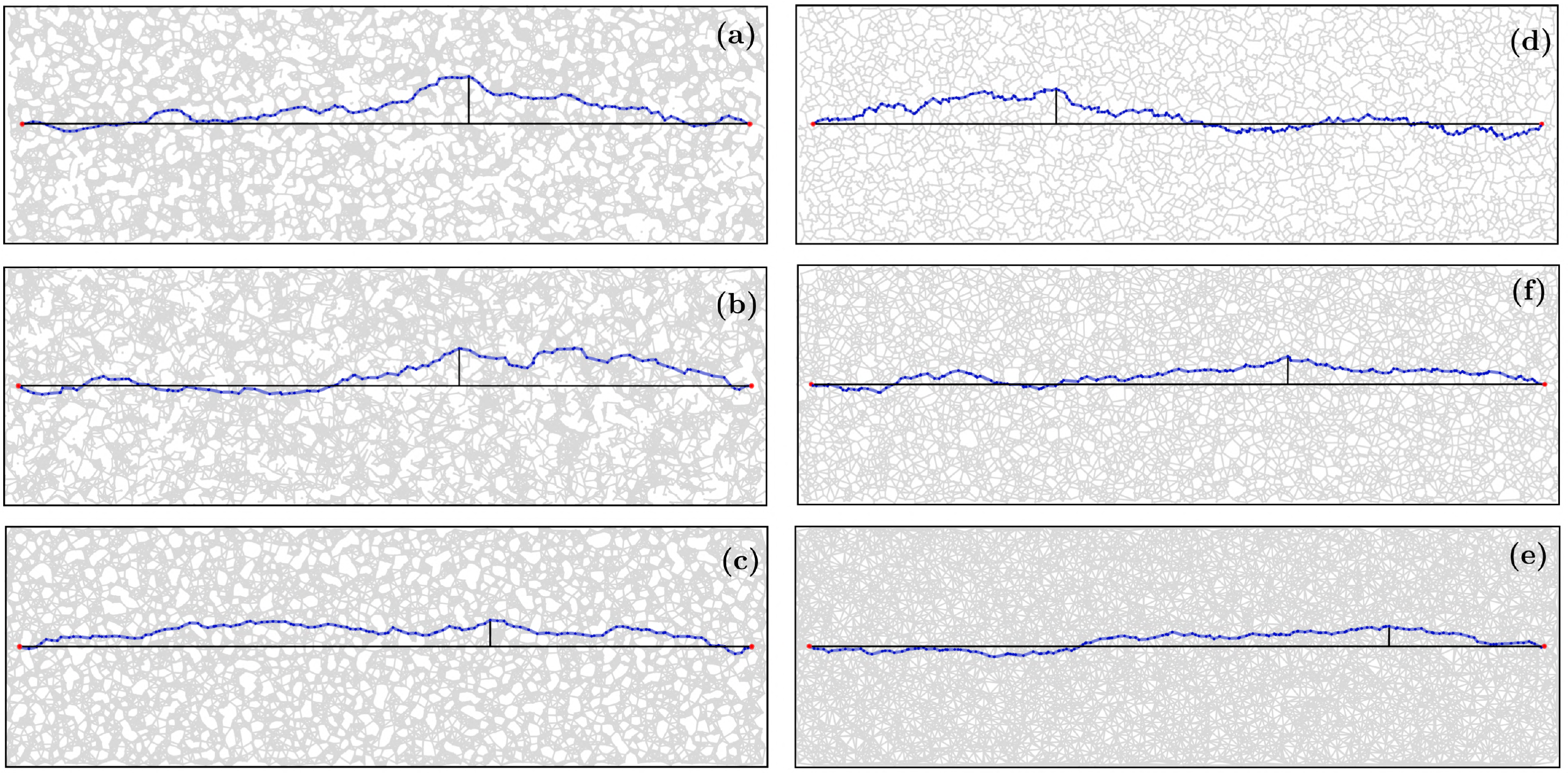}
     \end{adjustbox}
     \caption{Example Euclidean geodesics (blue) running between
       two end nodes of a simple, stationary Poisson point process (red). The maximal transversal
       deviation is shown (vertical black line). The Euclidean distance between the endpoints is the horizontal black line. The PPP density is equal for each model. (a) Hard RGG, (b) Soft RGG with connectivity probability $H(r)=\exp(-r^2)$, (c) $7$-NNG, (d) RNG, (e) GG, (f) DT.}
 \label{fig:diagram}
\end{figure*}

Given i.i.d weights, paths are sums of i.i.d. random variables. The
lengths of paths between pairs of points can be considered to be a
random perturbation of the plane metric. In fact, these lengths, and
the corresponding transversal deviations of the geodesics, have been
the focus of in-depth research over the last 50 years
\cite{auffinger2017}. They exist as minima over collections of
correlated random variables. The travel times are conjectured (in the
i.i.d.) case to converge to the Tracy-Widom distribution (TW), found
throughout various models of statistical physics, see
e.g. \cite[Section 1]{santalla2015}. This links the model to random
matrix theory, where $\beta$-TW appears as the limiting distribution
of the largest eigenvalue of a random matrix in the $\beta$-Hermite
ensemble, where the parameter $\beta$ is 1,2 or 4 \cite{keating2000}.

The original FPP model is defined as follows. We consider vertices in
the $d$-dimensional lattice $\mathbb{L}^{d} = (\mathbb{Z}^{d},E^{d})$
where $E^{d}$ is the set of edges. We then construct the function
$\tau: E^{d} \to (0,\infty)$ which gives a weight for each edge and
are usually assumed to be identically independently distributed random
variables. The passage time from vertices $x$ to $y$ is the random
variable given as the minimum of the sum of the $\tau$'s over all
possible paths $P$ on the lattice connecting these points,
\begin{align}
T(x,y)=\min_{P}\sum_P\tau(e)
\end{align}
This minimum path is a \textit{geodesic}, and it is
almost surely unique when the edge weights are continuous.

The average travel time is proportional to the distance
\begin{align}
  \mathbb{E}\left(T\left(x,y\right)\right)\sim |x-y|
\end{align}
where here and in the following we denote the average of a quantity by
$\mathbb{E}\left(\cdot\right)$, and where $a \sim b$ means $a$
converges to $Cb$ with $C$ a constant independent of $x,y$, as
$|x-y| \to \infty$. More generally, if the ratio of the geodesic length and the Euclidean
distance is less than a finite number $t$ (the maximum value of this
ratio is called the stretch), the network is a $t$\textit{-spanner}
\cite{narasimhan2007}. Many important networks are $t$-spanners,
including the Delaunay triangulation of a Poisson point process, which
has $\pi/2< t < 1.998$ \cite{bose2011,xia2011}. The variance of the
passage time over some 
distance $|x-y|$ is also important, and scales as
\begin{align}
  \text{Var}\left(T\left(x,y\right)\right) \sim |x-y|^{2\chi},
\end{align}
The maximum deviation $D(x,y)$ of the geodesic from the straight line from $x$
to $y$ is characterised by the wandering exponent $\xi$, i.e.
\begin{align}
\mathbb{E}(D(x,y))\sim |x-y|^{\xi}
\end{align}
for large $|x-y|$.
Knowing $\xi$ informs us about the geometry of geodesics between two
distant points. See Fig. \ref{fig:diagram} for an illustration of
wandering on different networks.

The other exponent, $\chi$, informs us about the variance of their
random length. Another way to see this exponent is to consider a ball
of radius $R$ around any point. For large $R$, the ball has an average
radius proportional to $R$ and the fluctuations around this average
grow as $R^\chi$ \cite{santalla2015}. With $\chi<1$ the fluctuations
die away $R \to \infty$, leading to the \textit{shape theorem}, see
e.g. \cite[Section 1]{auffinger2017}.

\subsubsection{Sublinear variance in FPP}

According to Benjamini, Kalai and Schramm,
$\text{Var}\left(T(x,y)\right)$ grows sub-linearly with $|x-y|$
\cite{benjamini2002}, a major theoretical step in characterising their
scaling behaviour. With $C$ some constant which depends only on the
distribution of edge weights and the dimension $d$, they prove that
\begin{equation}
  \text{Var}\left(T\left(x,y\right)\right) \leq C |x-y|/\log|x-y|.
\end{equation}
The numerical evidence, in fact, shows this follows the non-typical
scaling law $|x-y|^{2/3}$. Transversal fluctuations also scale as
$|x-y|^{2/3}$ \cite{auffinger2017}. In this case, the fluctuations of $T$ are
asymptotic to the TW distribution. According to recent results of
Santalla et al. \cite{santalla2017}, curved spaces lead to similar
fluctuations of a subtly different kind: if we embed the graph on the
surface of a cylinder, the distribution changes from the largest
eigenvalue of the GUE, to GOE, ensembles of random matrix theory.

When we see a sum of random variables, it is natural to
conjecture a central limit theorem, where the fluctuations of the sum, after
rescaling, converge to the standard normal distribution in some limit,
in this case as $|x-y|\to\infty$. Durrett writes
in a review that
\textit{``...novice readers would expect a central limit theorem being
  proved,...however physicists tell us that in two dimensions, the
  standard deviation is of order $|x-y|^{1/3}$''}, see \cite[Section
1]{benjamini2002}. This suggests that one does not have a Gaussian central limit theorem for the travel time fluctuations in the usual way. This has been rigourously proven
\cite{johansson2000,baik1999,deuschel1999}.

\subsubsection{Scaling exponents}

A well-known result in the 2d lattice case \cite{kardar1986} is that
$\chi=1/3$, $\xi=2/3$. Also, another belief is that $\chi=0$ for
dimensions $d$ large enough. Many physicists, see for example
\cite{huse1985,kardar1986,kardar1987,krug1987,krug1989,medina1989,krug1991},
also conjecture that independently from the dimension, one should have
the so-called KPZ relation between these exponents
\begin{equation}\label{e:kpz}
\chi = 2\xi - 1
\end{equation}
This is connected with the KPZ universality class of random geometry, apparent in many physical situations, including traffic and data flows, and their respective models, including the
corner growth model, ASEP, TASEP, etc \cite{deift2006,grimmettbook,derrida1998}. In
particular, FPP is in direct correspondence with important problems in
statistical physics \cite{halpin1995} such as the directed polymer in random media
(DPRM) and the KPZ equation, in which case the dynamical exponent $z$
corresponds to the wandering exponent $\xi$ defined for the FPP
\cite{calabrese2011,santalla2015}.

\subsubsection{Bounds on the exponents}

The situation regarding exact results is more complex \cite{chatterjee2013,auffinger2017}. The majority of results are based on the model on $\mathbb{Z}^d$. Kesten
\cite{kesten1993} proved that $\chi\leq 1/2$ in any dimension, and for
the wandering exponent $\xi$, Licea et al. \cite{licea1996} gave some
hints that possibly $\xi\geq 1/2$ in any dimension and $\xi\geq 3/5$ for $d=2$.

Concerning the KPZ relation, Wehr and Aizenman \cite{wehr1990} and Licea et al
\cite{licea1996} proved the inequality \begin{equation}
  \chi\geq (1-d\xi)/2\end{equation} in $d$
dimensions. Newman and Piza \cite{newman1995} gave some hints that possibly 
$\chi\geq 2\xi-1$. Finally,
Chatterjee \cite{chatterjee2013} proved Eq. \ref{e:kpz} for
$\mathbb{Z}^d$ with independent and identically distributed random
edge weights, with some restrictions on distributional properties of the weights. These
were lifted by independent work of Auffinger and Damron \cite{auffinger2017}.

\subsection{Euclidean first passage percolation}\label{sec:efpp}

Euclidean first passage percolation \cite{howard1997}
adopts a very different perspective from FPP by considering a fluid flowing along each of
the links of the complete graph on the points at some weighted speed
given by a function, usually a power, of the Euclidean length of the
edge. We ask, between two points of the
process at large Euclidean distance $\norm{x-y}$, what is the minimum passage time
over all possible routes.

More precisely, the original model of Howard and Newman goes as
follows. Given a domain $\mathcal{V}$ such as the Euclidean plane, and
$\mathrm{d}x$ Lesbegue measure on $\mathcal{V}$, consider a Poisson
point process $\mathcal{X} \subset \mathcal{V}$ of intensity
$\rho \mathrm{d}x$, and the function
$\phi: \mathbb{R}^{+} \to \mathbb{R}^{+}$ satisfying $\phi(0)=0$,
$\phi(1)=1$, and strict convexity. We denote by $K_{\mathcal{X}}$ the
complete graph on $\mathcal{X}$. We assign to edges $e = \{q,q'\}$
connecting points $q$ and $q'$ the weights $\tau(e) = \phi(\norm{q-q'})$, and a natural choice is
\begin{equation}
  \phi(x) = x^{\alpha}, \quad \alpha > 1
  \label{eq:phi}
\end{equation}
The reason for $\alpha>1$ is that the shortest path is otherwise the direct link, so this introduces non-trivial geodesics.

The first work on a Euclidean model of FPP concerned the Poisson-Voronoi tessellation of the $d$-dimensional
Euclidean space by Vahidi-Asl and Wierman in 1992
\cite{vahidiasl1992}. This sort of generalisation is a long term goal
of FPP \cite{auffinger2017}.  Much like the lattice model with
i.i.d. weights, the model is rotationally invariant. The corresponding
\textit{shape theorem}, discussed in \cite[Section 1]{auffinger2017},
leads to a ball. The existence of bigeodesics (two paths,
concatenated, which extend infinitely in two distinct directions from
the origin, with the geodesic between the endpoints remaining
unchanged), the linear rate of the local growth dynamics (the wetted
region grows linearly with time), and the transversal fluctuations of
the random path or surface are all summarised in \cite{auffinger2014}.

\subsubsection{Bounds on the exponents}

Licea et al \cite{licea1996} showed that for the standard
first-passage percolation 
on $\mathbb{Z}^d$ with $d\geq 2$, the wandering exponent satisfies
$ \xi(d)\geq 1/2$
and specifically
\begin{align}
\xi(2)\geq 3/5
\end{align}
In Euclidean
FPP, however, these bounds do \textit{not} hold, and we have \cite{howard2000,howard2001}
\begin{align}
\frac{1}{d+1}\leq \xi\leq 3/4
\end{align}
and, for the wandering exponent,
\begin{align}
&\chi\geq \frac{1-(d-1)\xi}{2}.
\end{align}
Combining these different results then yields, for $d=2$
\begin{align}
&1/8\leq \chi\\
&1/3\leq\xi\leq 3/4
\end{align}
Since the KPZ relation of Eq. \ref{e:kpz} apparently holds in our setting, the lower
bound for $\chi$ implies then a better bound for $\xi$, namely
\begin{align}
\xi\geq \frac{3}{3+d}
\end{align}
which in the two dimensional case leads to $\xi\geq 3/5$, the same
result as in the standard FPP.

Also, the `rotational invariance' of the Poisson point process implies the KPZ relation Eq. \ref{e:kpz} is satisfied in each spatial network we study. We numerically corroborate this in Section \ref{sec:results}. See for example \cite[Section 4.3]{auffinger2017} for a discussion of the generality of the relation, and the notion of rotational invariance.


\subsection{EFPP on a spatial network}\label{sec:efpp}

This is the model that we are considering here. Instead of taking as
in the usual EFPP into
account all possible edges with an exponent $\alpha>1$ in
Eq.~\ref{eq:phi}, we allow only some edges between the points and
take the weight proportional to their length (ie. $\alpha=1$ here). This leads to a different
model, but apparently universal properties of the geodesics. We
therefore move beyond the weighted complete graph of Howard and
Newman, and consider a large class of spatial networks, including the
random geometric graph (RGG), the $k$-nearest neighbour graph (NNG),
the $\beta$-skeleton (BS), and the Delaunay triangulation (DT). We
introduce them in Section \ref{sec:spanet}.


\section{Random spatial networks}
\label{sec:spanet}

We consider in this study spatial networks constructed over a set of
random points. We focus
on the most straightforward case, and consider a stationary Poisson
point process in the $d$-dimensional Euclidean space, taking
$d=2$. This constitutes a Poisson random number of points, with
expectation given by $\rho$ per unit area, distributed uniformly at
random. We do not discuss here typical generalisations, such as to the
Gibbs process, or Papangelou intensities \cite{last2017}.

First, we will consider the complete graph as in the usual EFPP, with edges weighted according to the details of Sec. \ref{sec:efpp}.
We will then consider the four distinct \textit{excluded region graphs} defined
below. Note that some of these networks actually obey inclusion
relations, see for example \cite{aldous2010}. We have
\begin{align}
RNG \subset GG \subset DT
\end{align}
where RNG stands for the relative neighborhood graph, GG for the Gabriel
graph, and DT for the Delaunay triangulation. This nested relation
trivially implies the following inequality
\begin{align}
\xi_{RNG}\geq \xi_{GG}\geq \xi_{DT}
\end{align}
as adding links can only decrease the wandering exponent. We are not
aware of a similar relation for $\chi$. We will also consider three
distinct \textit{proximity graphs} such as the hard and soft RGG, and
the $k$-nearest neighbour graph.

\subsection{Proximity graphs}

The main idea for constructing these graphs is that two nodes
have to be sufficiently near in order to be connected. 

\subsubsection{Random geometric graph}\label{sec:rgg}

The usual random geometric graph is defined in \cite{penrosebook}
and was introduced by Gilbert \cite{gilbert1961} who assumes
that points are randomly located in the plane and
have each a communication range $r$. Two nodes are connected by an edge if they
are separated by a distance less than $r$.

We also have the following variant: the \textit{soft} random geometric graph \cite{coon2012,penrose2016,kartungiles2018}. This is a graph formed
on $\mathcal{X} \subset \mathbb{R}^{d}$ by adding an edge between
distinct pairs of $\mathcal{X}$ with probability $H(|x-y|)$ where
$H: \mathbb{R}^{+} \to \left[0,1\right]$ is called the
\textit{connection function}, and $|x-y|$ is Euclidean distance.

We focus on the case of \textit{Rayleigh
fading} where, with $\gamma > 0$ a parameter and $\eta > 0$ the path
loss exponent, the connection function, with $|x-y|>0$, is given
by
\begin{eqnarray}\label{e:1}
H(\norm{x-y}) = \exp\left(-\gamma\norm{x-y}^{\eta}\right)
\end{eqnarray}
and is otherwise zero. This choice is discussed in
\cite[Section 2.3]{giles2016}.

This graph is connected with high probability when the mean degree is
proportional to the logarithm of the number of nodes in the graph. For
the hard RGG, this is given by $\rho \pi r^2$, and otherwise the
integral of the connectivity function over the region visible to a
node in the domain, scaled by $\rho$ \cite{penrose2016}. As such, the
graph must have a very large typical degree to connect.

\subsubsection{$k$-Nearest Neighbour Graph}

For this graph, we connect points to their
$k \in \mathbb{N}$ nearest neighbours. When $k=1$, we obtain the
nearest neighbour graph ($1$-NNG), see e.g. \cite[Section
3]{waltersreview}. The model is notably different from the RGG because
local fluctuations in the density of nodes do not lead to local
fluctuations in the degrees. The typical degree is much lower than the
RGG when connected \cite{waltersreview}, though still remains
disconnected on a random, countably infinite subset of the
$d$-dimensional Euclidean space, since isolated subgraph exist. For
large enough $k$, the graph contains the RGG as a subgraph. See
Section \ref{sec:exponents} for further discussion.

\subsection{Excluded region graphs}

\begin{figure}[!]
       \begin{adjustbox}{width=\linewidth}
    \includegraphics[scale=0.36]{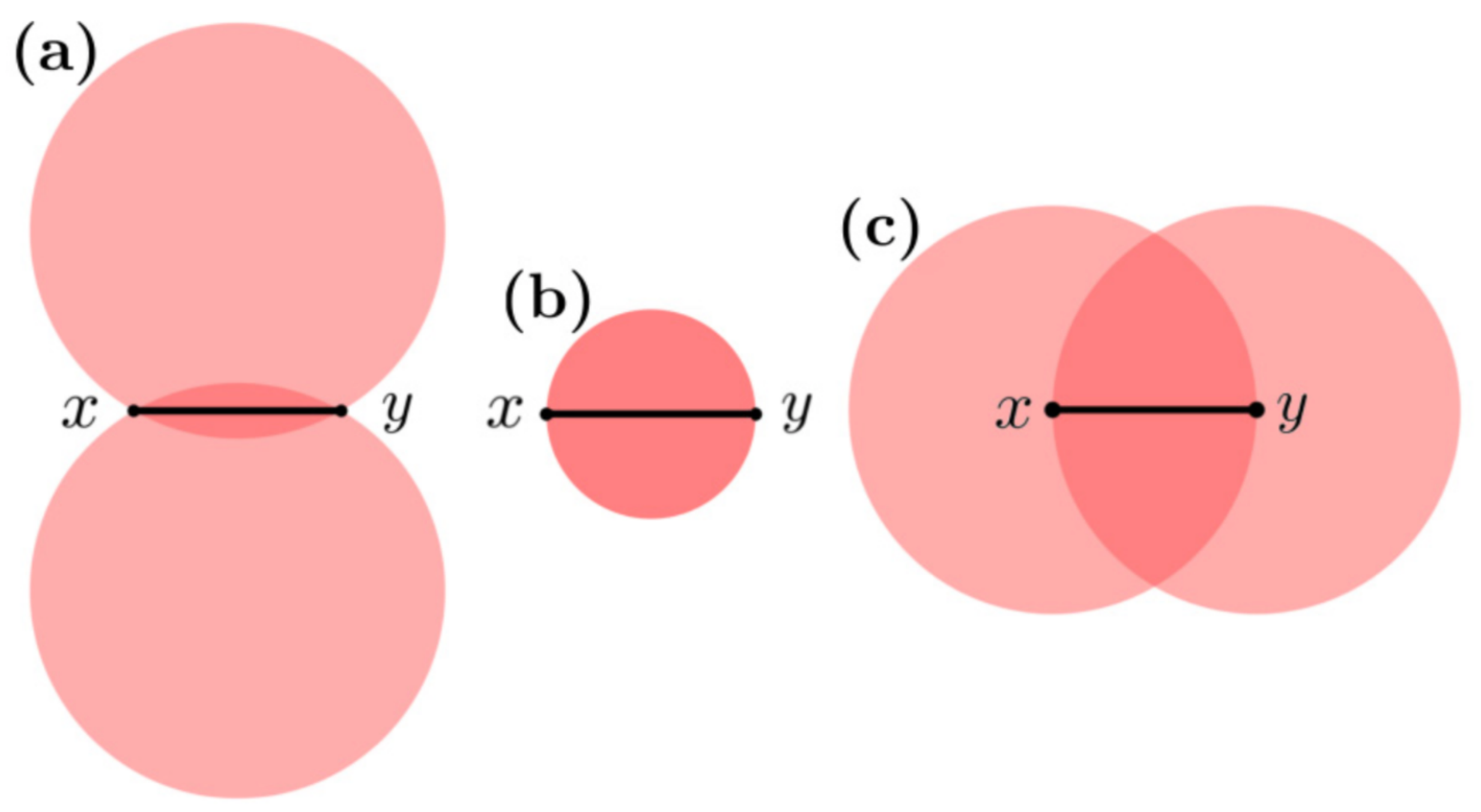}
    \end{adjustbox}
    \caption{The geometry of the lune-based $\beta$-skeleton for (a)
      $\beta=1/2$, (b) $\beta=1$, and (c) $\beta=2$. For $\beta<1$, nodes within the
      intersection of two disks each of radius $|x-y|/2\beta$
      preclude the edges between the disk centers, whereas for
      $\beta>1$, we use instead radii of $\beta |x-y|/2$. Thus,
      whenever two nodes are nearer each other than any other
      surrounding points, they connect, and otherwise do
      not.}\label{fig:betapic}
\end{figure}

The main idea here for constructing these graphs is that two nodes
will be connected if some region between them is empty of points. See Fig. \ref{fig:betapic} for a depiction
of the geometry of the lens regions for $\beta-$skeletons.

\subsubsection{Delaunay triangulation}

The Delaunay triangulation of a set of points is the dual graph of
their Voronoi tessellation. One builds the graph by trying to match
disks to pairs of points, sitting just on the perimeter, without
capturing other points of the process within their bulk. If and only
if this can be done, those points are joined by an edge. The
triangular distance Delaunay graph can be similarly constructed with a
triangle, rather than a disk, but we expect universal exponents.

For each simplex within the convex hull of the
triangulation, the minimum angle is maximised, leading in general to
more realistic graphs. It is also a
\textit{t-spanner} \cite{narasimhan2007}, such that with $d=2$ we have the geodesic between
two points of the plane along edges of the triangulation to be no more
than $t<1.998$ times the Euclidean separation \cite{xia2011}. The DT is necessarily connected.

\subsubsection{$\beta$-skeleton}

The lune-based $\beta$-skeleton is a way of naturally capturing the shape of
points \cite[Chapter 9]{deberg2008}. see Fig. \ref{fig:betapic}.

A \textit{lune} is the intersection
of two disks of equal radius, and has a \textit{midline} joining the centres of the disks and two corners on its perpendicular
bisector.  For $\beta \leq 1$, we define the excluded region of each pair of points $(x,y)$ to be the lune of radius $|x-y| / 2 \beta$
with corners at $x$ and $y$.  For $\beta \geq 1$ we use instead the lune of radius $\beta|x-y|/2$, with $x$ and $y$ on the midline.  For each
value of beta we construct an edge between each pair of points if and only if its excluded region is empty.  For
$\beta=1$, the excluded region is a disk and the beta-skeleton is called the \textit{Gabriel graph} (GG), whilst for $\beta=2$ we have the \textit{relative neighbourhood graph} (RNG).

For $\beta \leq 2$, the graph is necessarily connected. Otherwise, it is typically disconnected.







\section{Numerical results}
\label{sec:results}
\subsection{Numerical setup}
Given the models in the previous section, we numerically evaluate the
scaling exponents $\chi$ and $\xi$, as well as the distribution of the
travel time fluctuations. We now describe the numerical setup. With
density of points $\rho>0$, and a small tolerance $\epsilon$, we
consider the rectangle domain
\begin{equation}
  \mathcal{V} = [-w/2-\epsilon/2,w/2+\epsilon/2] \times [-h/2,h/2],
\end{equation}
and place a
\begin{equation}
 n \sim \text{Pois}\left(h\left(w+\epsilon\right) \rho \right)
\end{equation}
points uniformly at random in
$\mathcal{V}$. Then, on these random points, we build a spatial network by connecting pairs of points according to the rules of either the NNG, RGG, $\beta$-skeleton for $\beta=1,2$, the DT, or the weighted complete graph of EFPP.

Two extra points are fixed near the boundary arcs at $(-w/2,0)$ and
$(w/2,0)$, and the Euclidean geodesic is then identified using a
variant of Djikstra's algorithm, implemented in Mathematica 11. The
tolerance $\epsilon$ is important for the Soft RGG, since this graph
can display geodesics which reach backwards from their starting point,
or beyond their destination, before hopping back. We set
$\epsilon = w/10$. This process is repeated for $N=2000$
graphs, each time taking only a single sample of the geodesic length
over the span $w$ between the fixed points on the boundary. This act
of taking only a single path is done to avoid any small correlations
between their statistics, since the exponents are vulnerable to tiny
errors given we need multiple significant figures of precision to draw
fair conclusions. It also allows us to use smaller domains. The
relatively small value for $N$ is sufficient to determine the
exponents at the appropriate computational speed for the larger
graphs.

The approach in \cite{santalla2015} involves rotating the point
process before each sample is taken, which is valid alternative
method, but we, instead, aim for maximium accuracy given the exponents
are not previously conjectured, and therefore need to be determined
with exceptional sensitivity, rather than at speed. Note that the fits
that we are doing here are over the same typical range as in this work  \cite{santalla2015}.

We then increase $w$, in steps of 3 units of distance, and repeat,
until we have statistics of all $w$, up to the limit of computational
feasibility. This varies slightly between models. The RGGs are more
difficult to simulate due to their known connectivity constraint
where vertex degrees must approach infinity, see e.g. \cite[Chapter
1]{penrosebook}. Thus we cannot simulate connected graphs to the same
limits of Euclidean span as with the other models.

We are then able to relate the mean and standard deviation of the
passage time, as well as the mean wandering, to $w$, at various
$\rho$, and for each model. For example, the left hand plots in
Fig. \ref{fig:taball} show that the typical passage time
$\mathbb{E}T(x,y) \sim w$, i.e. grows linearly with $w$, for all
networks \cite{aldous2009,aldous2010,kartungiles2018}. The standard
error is shown, but is here not clearly distinguishable from the symbols.

 We ensure $h$ is large enough to stop the geodesics hitting the boundary, so
we use a domain of height equal to the mean deviation
$\mathbb{E}D(w)$, plus six standard deviations.

The key computational
difficulty here is the memory requirement for large graphs, of which
all $N$ are stored simultaneously, and mapped in parallel on a Linux
cluster over a function which measures the path statistic. This
parallel processing is used to speed up the computation of the geodesics lengths and wandering.

\subsection{Scaling exponents}

\begin{figure*}[ht!]
  \begin{adjustbox}{width=\linewidth}
    \includegraphics[scale=0.36]{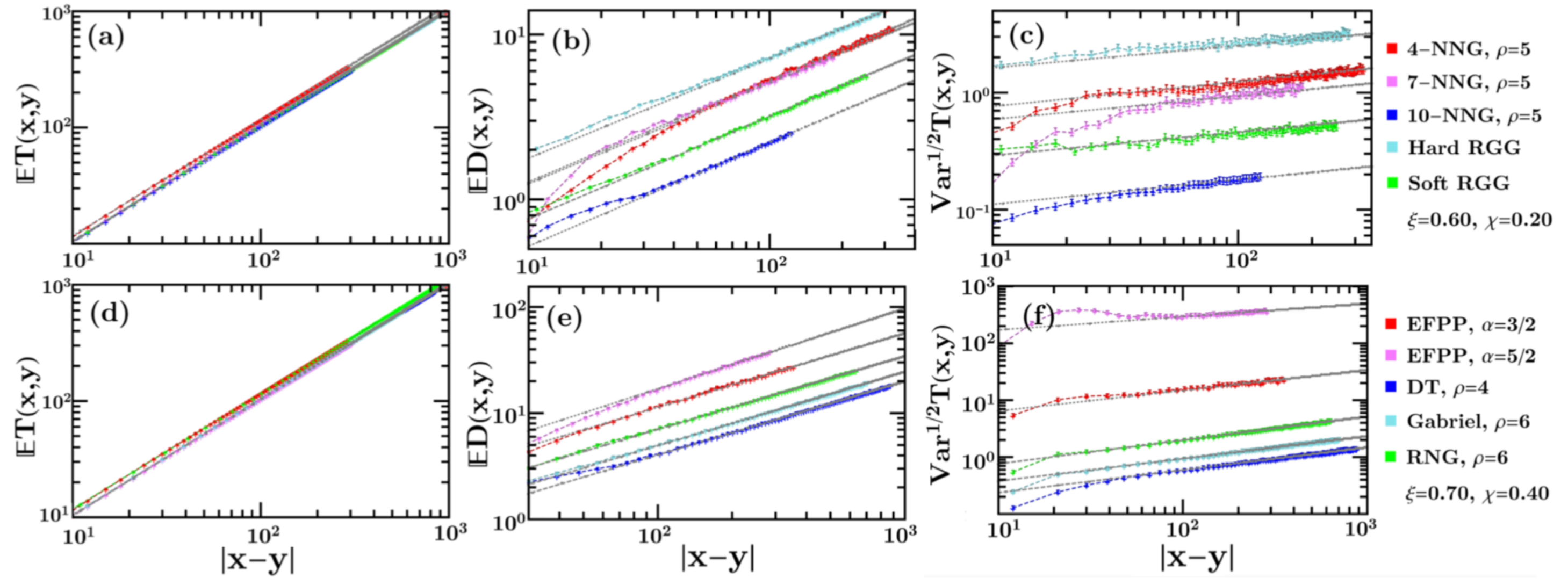}
   \end{adjustbox}
   \caption{The three statistics we observe, expected travel time
     (a) and (d), expected wandering (b) and (e), and standard deviation of the travel time (c) and (f). The power law exponents are indicated in the
     legend. Error bars of one standard deviation are shown for each
     point. The top plots show the results from the models in first
     universality class, while the lower plots show the second
     class. The RGG and NNG are distinguished with different colours
     (green and blue), as are EFPP on the complete graph, the DT, and the two beta skeletons
     (Gabriel graph, and relative neighbourhood graph). The point process density $\rho$ points per unit area is given for each model.}
 \label{fig:taball}
\end{figure*}

The results are shown in Figs.~\ref{fig:taball}. These plots, shown in
loglog, reveal a power law behaviour of $T$ and $D$, and the linear
growth of typical travel time with Euclidean span. We then compute the
exponents to two significant figures using a nonlinear model fit,
based on the model $a |x-y|^{b}$, and then determining the
parameters $a,b$ using the quasi-Newton method in Mathematica 11.

Our numerical results suggest that we can distinguish two classes of spatial network models by
the scaling exponents of their Euclidean geodesics. The proximity
graphs (hard and soft RGG, and $k$-NNG) are in one class, with
exponents
\begin{align}
  &\chi_{RGG,NNG}=0.20 \pm 0.01 \\
  &\xi_{RGG,NNG}=0.60 \pm 0.01
\end{align}
whereas the excluded region graphs (the $\beta$-skeletons and Delaunay
triangulation), and Howard's EFPP model with $\alpha>1$, are in another class with
\begin{align}
  &\chi_{DT,\beta \text{-skel},\text{EFPP}}=0.40 \pm 0.01 \\
  &\xi_{DT,\beta \text{-skel},\text{EFPP}}=0.70 \pm 0.01
\end{align}

Clearly, the KPZ relation of Eq. \ref{e:kpz} is satisfied up to the
numerical accuracy which we are able to achieve. We corroborate that
this is independent of the density of points and connection range
scaling, given the graphs are connected. The exponents hold
asymptotically i.e. large inter-point distances. Thus we conjecture
\begin{align}
  \text{Var}\left(T\left(x,y\right)\right) \sim |x-y|^{4/5},\\
  \mathbb{E}\left(D\left(x,y\right)\right) \sim |x-y|^{7/10}
\end{align}
for the proximity graphs (the DT and the $\beta$-skeletons for all
$\beta$), and, for the RGGs and the $k$-NNG,
\begin{align}
  \text{Var}\left(T\left(x,y\right)\right) \sim |x-y|^{2/5},\\
  \mathbb{E}\left(D\left(x,y\right)\right) \sim |x-y|^{3/5},
\end{align}
We summarize these new results in Table \ref{table:exp}.
\begin{table}
\caption{Exponents $\xi$ and $\chi$, and passage time distribution for
  the various networks considered.}
\centering
\begin{tabular}{c c c c} 
\hline\hline 
Network & $\xi$ & $\chi$ & Distribution of $T$\\ [0.5ex] 
\hline 
\textit{Proximity graphs} & & &\\
Hard RGG & 3/5 & 1/5 & Normal (Conj.)\\
Soft RGG with Rayleigh fading & 3/5 & 1/5 & Normal (Conj.)\\
$k$-NNG & 3/5 & 1/5 & Normal\\
\hline 
\textit{Excluded region graphs} &  &  &\\
DT & 7/10 & 2/5  & Normal\\
GG & 7/10 & 2/5 & Normal\\
$\beta-$skeletons & 7/10 & 2/5 & Normal\\
RNG & 7/10 & 2/5 & Normal\\ [1ex] 
\hline 
\textit{Euclidean FPP} &  &  &\\
With $\alpha=3/2$ & 7/10 & 2/5  & Normal\\
With $\alpha=5/2$ & 7/10 & 2/5 & Normal\\[1ex] 
\hline 
\end{tabular}
\label{table:exp} 
\end{table}
It is surprising that these exponents are apparently rational numbers. In Bernoulli
continuum percolation, for example, the threshold connection range for
percolation is not known, but not thought to be rational, as it is
with bond percolation on the integer lattice \cite[Chapter
10]{penrosebook}. Exact exponents are not necessarily expected in the
continuum setting of this problem, which suggests there is more to be
said about the classification of first passage process via this
method.

\subsection{Travel time fluctuations}\label{sec:traveltime}

\begin{figure*}[ht!]
  \begin{adjustbox}{width=\linewidth}
    \centering
    \includegraphics[scale=0.4]{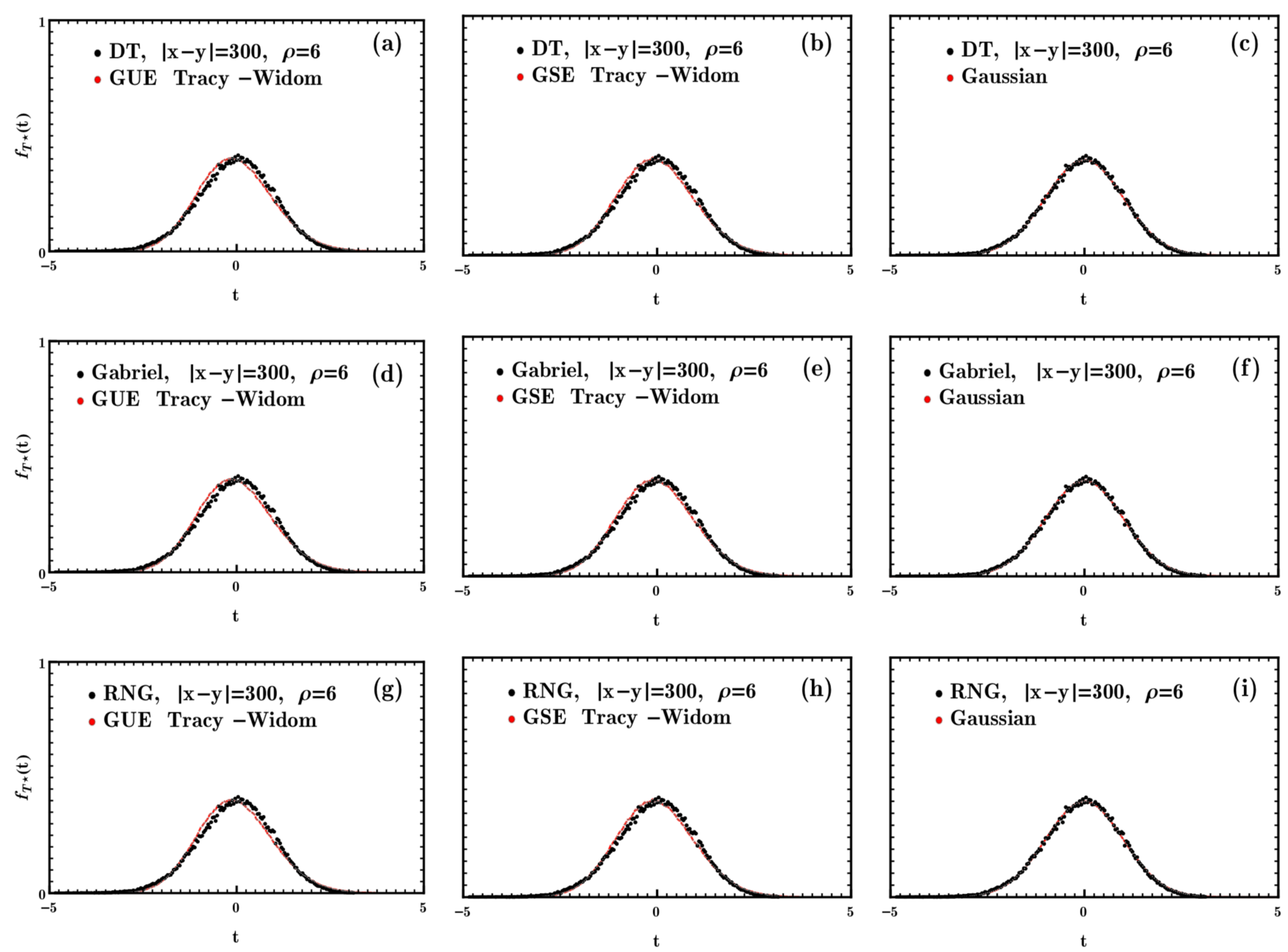}
      \end{adjustbox}

      \caption{Travel time distributions for the DT (a)-(c), RNG (d)-(f), and Gabriel (g)-(i)
        graphs, compared with the GUE and GSE Tracy-Widom ensembles,
        and the Gaussian distribution. The point process density $\rho$ points per unit area is given for each model. The slight skew of the TW
        distribution is not present in the data.}
   \label{fig:rggdist}
\end{figure*}

\begin{figure*}
    \centering
         \begin{adjustbox}{width=\linewidth}
    \centering
    \includegraphics[scale=0.19]{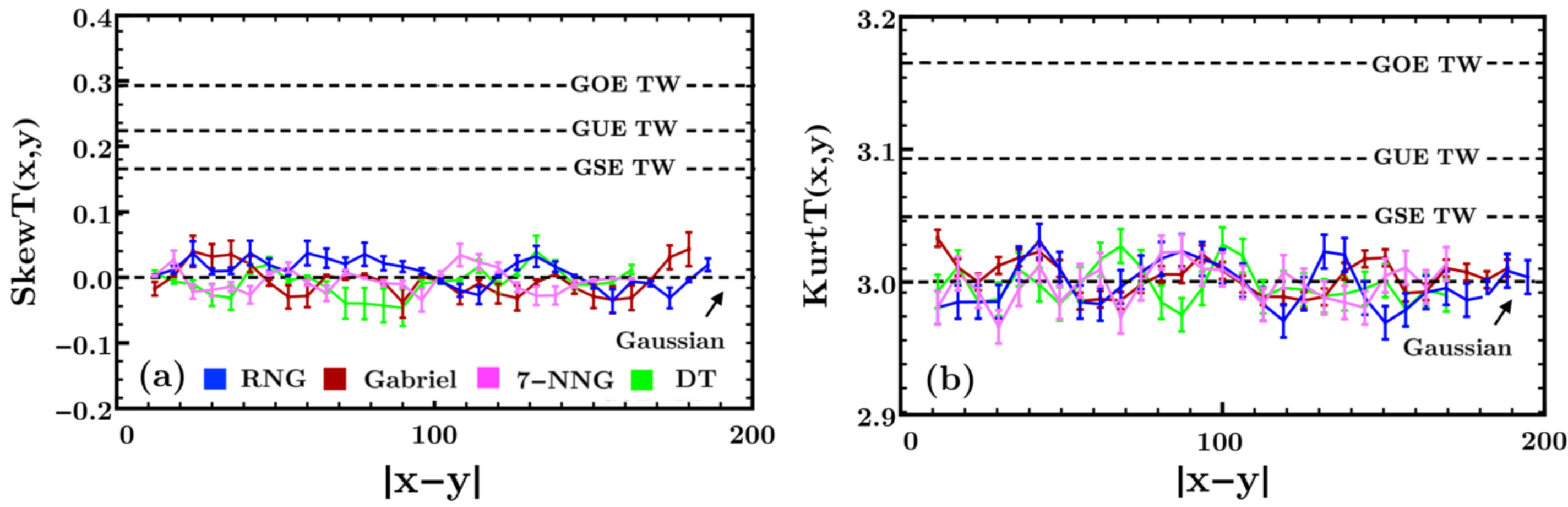}
         \end{adjustbox}

         \caption{Skewness (a) and kurtosis (b), for the travel time
           fluctuations, computed for each network model. For a
           Gaussian distribution, the skewness is $0$ and the kurtosis
           equal to $3$, values that we indicate by dashed black lines. The point process density $\rho$ points per unit area is given for each model. The
           Tracy-Widom distribution has only marginally different moments to the normal, also shown by dashed black lines, with labels added to distinguish each specific distribution (GOE, GUE or GSE), as well as the Gaussian. }\label{fig:moments}
\end{figure*}

We see numerically that the travel time distribution is a normal for
most cases (see Fig. \ref{fig:rggdist}). We summarise these results in the Table \ref{table:exp}
and in Fig. \ref{fig:moments} we show the skewness and kurtosis for
the travel time fluctuations, computed for the different networks. For
a Gaussian distribution, the skewness is $0$ and the kurtosis equal to
$3$, while the Tracy-Widom distribution displays other values.

We provide some detail of the distribution of $T$ for each model from the proximity class in
Fig. \ref{fig:rggdist}. This is compared
against four test distributions, the Gaussian orthogonal, unitary and
symplectic Tracy Widom distributions, and also
the standard normal distribution.

This makes the case of EFPP on spatial networks one of only a
few special cases where Gaussian fluctuations in fact occur. Auffinger
and Damron go into detail concerning each of the remaining cases in
\cite[Section 3.7]{auffinger2017}. One example, reviewed extensively
by Chaterjee and Dey \cite{chatterjee2013}, is when geodesics are
constrained to lie within thin cylinders i.e. ignore paths which
traverse too far, and thus select the minima from a subdomain. This
result could shed some light on their questions, though in what way it
is not clear.

We also highlight that Tracy-Widom is thought to occur in problems
where matrices represent collections of totally uncorrelated random
variables \cite{mopost}. In the case of EFPP, we have the interpoint
distances of a point process, which lead to spatially correlated
interpoint distances, so the adjacency matrix does not contain
i.i.d. values. This potentially leads to the
loss of Tracy-Widom. However, we also see some cases of $N \times N$
large complex correlated Wishart matrices leading to TW for at least
one of their eigenvalues, and with convergence at the scale $N^{2/3}$
\cite{hachem2016}.

\subsection{Transversal fluctuations}\label{sec:transversal}

\begin{figure}[ht!]
    \centering
                 \begin{adjustbox}{width=\linewidth}
    \centering
\includegraphics[scale=0.22]{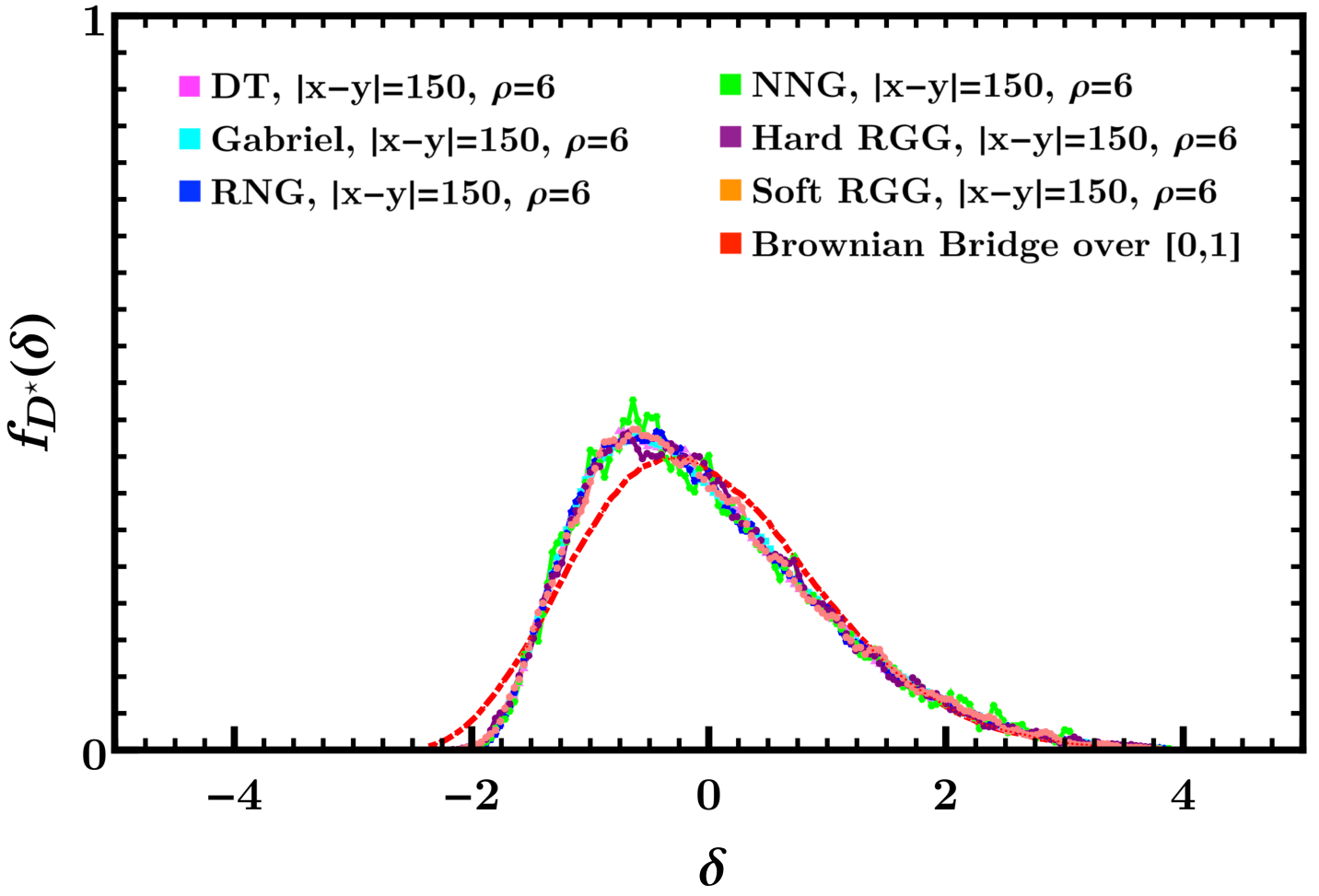} \hspace{1mm}
  \end{adjustbox}
  \caption{Transversal fluctuations of the geodesics in all models
    (coloured points), and compared with the fluctuations of a continuous
    Brownian bridge process between the same endpoints (red, dashed
    curve). The point process density $\rho$ points per unit area is given for each model.}\label{fig:devdist}
      \end{figure}

The transversal deviation distribution results appear beside
our evaluation of the scaling exponents, in
Fig. \ref{fig:devdist}. All the models produce geodesics with the same transversal fluctuation
distribution, despite distinct values of $\xi$. The fluctuations are
also distinct from the Brownian bridge (a geometric brownian motion
constrained to start and finish at two fixed position vectors in the
plane), running between the midpoints of the boundary arcs
\cite{grimmettbook}. It is a key open question to provide some
information about this distribution, as it is rarely studied in any
FPP model, as far as we are aware of the literature. A key work is
Kurt Johannson's, where the wandering exponent is derived analytically
in a variant of oriented first passage percolation. One might ask if a
similar variant of EFPP might be possible \cite{johansson2000}.

\section{Discussion}
\label{sec:5}

The main results of our investigation are the new rational exponents
$\chi$ and $\xi$ for the various spatial models, and the discovery of
the unusual Gaussian fluctuations of the travel time. We found that
for the different spatial networks the KPZ relation holds and known bounds are
satisfied. Also, due to known relations and the the KPZ law we have
\begin{align}
  \frac{3}{3+d}\leq \xi \leq \frac{3}{4}
\end{align}
It is surprising to find a large class of networks, in particular the
Delaunay triangulation, that displays an exponent $\xi = 7/10$ and
points to the question of the existence of another class of graphs
which display the theoretically maximal $\xi=3/4$.

Both immediately present a number of open questions and topics of
further research which may shed light on the first passage process on
spatial networks. We list below a number of questions that we think
are important. 

\subsection{Gaussian travel time fluctuations}\label{sec:exponents}

We are not able to conclude that all the models in the proximity graph
class $\chi=3/5$, $\xi=1/5$ have Gaussian fluctuations in the travel
time. This is for a technical reason. All the models we
study are either connected with probability one, such as the DT or
$\beta$-skeleton with $\beta \leq 2$, or have a connection probability
which goes to one in some limit. We require connected graphs, or paths
do not span the boundary arcs, and the exponents are not well defined.

Thus, the difficult models to simulate are the HRGG, SRGG and $k$-NNG,
since these are in fact disconnected with probability one without
infinite expected degrees i.e. the dense limit of Penrose, see
\cite[Chapter 1]{penrosebook}, or with the fixed degree of the $k$-NNG $k = \Theta(\log n)$ and
$n \to \infty$ in a domain with fixed density and infinite
volume. Otherwise we have isolated vertices, or isolated subgraphs,
respectively.

However, the $k$-NNG has typically shorter connection range i.e. in
terms of the longest edge, and shortest non-edge, where the `length of a non-edge' is the corresponding interpoint distance between the disconnected vertices \cite[Section
  3]{waltersreview}. So the computations used to produce these
graphs and then evaluate their statistical properties are
significantly less demanding. Thus, the HRGG is computationally
intractable in the necessary dense limit, so we are unable to verify
the fluctuations of either $T$ or $D$. However, we can see a skewness
and kurtosis for $T(|x-y|)$ which are monotonically decreasing with
$|x-y|$, towards the hypothesised limiting Gaussian statistics, at
least for the limited Euclidean span we can achieve.

Given the $k$-NNG is in the same class, we are left to conjecture
Gaussian fluctuations hold throughout all the spatial models described
in Section \ref{sec:spanet}. It remains an open question to identify
any exceptional models where this does not hold.

\subsection{Percolation and connectivity}\label{sec:exponents}

If we choose two points at a fixed Euclidean distance, then simulate a
Poisson point process in the rest of the $d$-dimensional plane, construct the relevant
graph, and consider the probability that both points are in the giant
component, this is effectively a positive constant for reasonable
distances, assuming that we are above the percolation transition.  At
small distances, the two events are positively correlated. Thus, one
can condition on this event, and therefore when simulating, discount
results where the Euclidean geodesic does not exist. This defines FPP
on the giant component of a random graph.

It's not clear from our experiments whether the rare isolated nodes,
or occasionally larger isolated clusters, either in the RGGs or
$k$-NNG, affect the exponents.  One similar model system would be the
Lorentz gas: put disks of constant radius in the plane, perhaps at
very low density, and seek the shortest path between two points that
does not intersect the disks. The exponents $\chi$ and $\xi$ for this
setting are not known \cite{dettmann2012,grimmettbook}.

An alternative to giant component FPP would be to condition on the two
points being connected to each other.  This would be almost identical
for the almost connected regime, but weird below the percolation
transition.  In that case the event we condition on would have a
probability decaying exponentially with distance, and the point
process would end up being extremely special for the path to even
exist.  For example, an extremely low density RGG would be almost
empty apart from a path of points connecting the end points, with a
minimum number of hops.

\subsection{Betweenness centrality}

The extent to which nodes take part in shortest paths throughout a
network is known as \textit{betweenness centrality}
\cite{barthelemy2011,kirkley2018}. We ask to what extent
knowledge of wandering can lead to understanding the centrality of
nodes. The variant \textit{node shortest path} betweenness centrality
is highest for nodes in bottlenecks. Can this centrality index be
analytically understood in terms of the power law scaling of $D$? Is
the exponent directly relevant to its large scale behaviour?

In order to illustrate more precisely this question, let $G$ be the graph formed on a point process $\mathcal{X}$ by joining pairs of
points with probability $H(|x-y|)$. Consider $\sigma_{xy}$ to be the number
of shortest paths in $G$ which join vertices $x$ and $y$ in $G$, and
$\sigma_{xy}(z)$ to be the number of shortest paths which join $x$ to
$y$ in $G$, but also run through $z$, then with $\neq$ indicating a
sum over unordered pairs of vertices not including $z$, define the
betweenness centrality $g(z)$ of some vertex $z$ in $G$ to be
\begin{equation}\label{e:bc}
  g(z)=\sum_{i \neq j \neq k} \frac{\sigma_{ij}(z)}{\sigma_{ij}}
\end{equation}

In the continuous limit for dense networks we can discuss the betweenness centrality and
we recall some of the results of \cite{kog105}. More
precisely, we define $\chi_{xy}(z)$ as the indicator which gives unity whenever $z$
intersects the shortest path connecting the $d$-dimensional positions
$x,y \in \mathcal{V}$. Then the normalised betweenness $g(z)$ is given by
\begin{equation}\label{e:bc}
  g(z) = \frac{1}{\int_{\mathcal{V}^{2}} \chi_{xy}(\mathbf{0})
    \mathrm{d}x
    \mathrm{d}y}\int_{\mathcal{V}^{2}} \chi_{xy}(z) \mathrm{d}x \mathrm{d}y
\end{equation}
Based on the assumption that there exists a single topological
geodesic as $\rho \to \infty$, and that this limit also results in an
infinitesimal wandering of the path from a straight line segment, an
infinite number of points of the process lying on this line segment
intersect the topological geodesic as $\rho \to \infty$, assuming the
graph remains connected, and so $\chi_{xy}(z)$ can then written
as a delta function of the transverse distance from $z$ to the
straight line from $x$ to $y$. The betweenness can then be computed and we obtain \cite{kog105}
(normalised by its maximum value at
$g(0)$)
\begin{equation}
g(\epsilon) =\frac{2}{\pi}\left(1-\epsilon^{2}\right)E\left(\epsilon\right)
\label{e:12}
\end{equation}
where $E\left(k\right)=\int_{0}^{\pi/2}d\theta\left(1-k^{2}\sin^{2}\left(\theta\right)\right)^{1/2}$
is the complete elliptic integral of the second kind. We have also rescaled such that $\epsilon$ is in units of $R$.

Take $D(x,y)$ to be the maximum deviation from the horizontal of the Euclidean geodesic. Numerical simulations suggest that
\begin{equation}\label{e:c1}
\mathbb{E}D(x,y) = C\norm{x-y}^{\xi} 
\end{equation}
where the expectation is taken over all point sets $\mathcal{X}$. The `thin cylinders' are given by a Heaviside Theta function, so
assume that the characteristic function is no longer a delta spike,
but a wider cylinder
\begin{equation}\label{e:c1}
 \chi_{xy}(z) = \theta\left(D(x,y)-z_{\perp}\right)
\end{equation}
where $z_{\perp}$ is the magnitude of the perpendicular deviation of the position $z$ from $\text{hull}(x,y)$. We then have that
\begin{equation}\label{e:bc}
  g(z) = \frac{1}{\int_{\mathcal{V}^{2}}
    \theta\left(D-\mathbf{0}_{\perp}\right)
    \mathrm{d}x \mathrm{d}y}\int_{\mathcal{V}^{2}} \theta\left(D-z_{\perp}\right) \mathrm{d}V
\end{equation}
(where $\mathbf{0}$ is the transverse vector computed for the
origin). This quantity is certainly difficult to estimate, but
provides a starting point for computing finite density corrections to
the result of \cite{kog105}.

The boundary of the domain is crucial
in varying the centrality, which is something we ignore here. Without
an enclosing boundary, such as with networks embedded into spheres or
tori, the typical centrality at a position in the domain is uniform,
since no point is clearly distinguishible from any other. This is
discussed in detail on \cite{kog105}. In fact, a significant amount of
recent work on random geometric networks has highlighted the
importance of the enclosing boundary \cite{giles2016,coon2012}.

\section{Conclusions}
\label{sec:6}

We have shown numerically that there are two distinct universality
classes in Euclidean first passage percolation on a large class of
spatial networks. These two classes correspond to the following two
broad classes of networks: firstly, based on joining vertices
according to critical proximity, such as in the RGG and the NNG, and
secondly, based on graphs which connect vertices based on excluded
regions, as in the lune-based $\beta$-skeletons, or the DT. Heuristically, the most
efficient way to connect two points is via the nearest neighbour, which
suggests that $\xi$ for proximity graphs should on the whole be
smaller than for exclusion-based graphs, which is in agreement with
our numerical observations.

The passage times show Gaussian fluctuations in all models, which we
are able to numerically verify. This is a clear distinction between EFPP and
FPP. After similar results of Chaterjee and Dey \cite{chatterjee2013},
it remains an open question why this happens, and also of course how
to rigorously prove it.

We also briefly discussed notions of the universality of betweenness
centrality in spatial networks, which is influenced by the
wandering of shortest paths. A number of open questions remain about the range of
possible universal exponents which could exist on spatial networks,
whose characterisation would shed light on the interplay
between the statistical physics of
random networks, and their spatial counterparts, in way which could reveal deep insights about universality and geometry more generally.

\begin{center}
  {\normalsize \bf{Acknowledgements}}
  \end{center}
The authors wish to thank  M\'{a}rton Bal\'{a}zs and  B\'{a}lint T\'{o}th for a number of very helpful discussions, as well as Ginestra Bianconi at QMUL, J\"{u}rgen Jost at MPI Leipzig, and the School of Mathematics at the University of Bristol, who provided generous hosting for APKG while carrying out various parts of this research. This work was supported by the EPSRC project ``Spatially Embedded Networks'' (grant EP/N002458/1). APKG was partly supported by the EPSRC project ``Random Walks on Random Geometric Networks'' (grant EP/N508767/1). All underlying data are reproduced in full within the paper.


\end{document}